\documentstyle[epsfig]{elsart}

\begin{document}

\begin{frontmatter}

\title{ \bf One-neutron removal reactions on neutron-rich psd-shell nuclei }

\author[LPC]{E.~Sauvan},
\author[LPC,IFIN]{F.~Carstoiu},
\author[LPC]{N.A.~Orr\thanksref{email}},
\author[LPC]{J.C.~Ang\'elique},
\author[Surrey]{W.N.~Catford},
\author[Bham]{N.M.~Clarke},
\author[GANIL]{M. Mac Cormick\thanksref{thIPN}},
\author[Surrey]{N.~Curtis\thanksref{thFSU}},
\author[Bham]{M.~Freer},
\author[IPN]{S.~Gr\'evy\thanksref{thLPC}},
\author[LPC]{C.~Le~Brun},
\author[GANIL]{M.~Lewitowicz},
\author[LPC]{E.~Li\'egard},
\author[LPC]{F.M.~Marqu\'es},
\author[GANIL]{P.~Roussel-Chomaz},
\author[GANIL]{M.G.~Saint~Laurent},
\author[Surrey]{M.~Shawcross},
\author[LPC]{J.S.~Winfield\thanksref{thINFN}}

\address[LPC]{Laboratoire de Physique Corpusculaire,
IN2P3-CNRS, ISMRA et Universit\'e de Caen, F-14050 Caen cedex, France}
\address[IFIN]{IFIN-HH, P.O.~Box MG-6, 76900 Bucharest-Magurele, Romania}
\address[Surrey]{Department of Physics, University of Surrey,
Guildford, Surrey, GU2~5XH, United Kingdom}
\address[Bham]{School of Physics and Astronomy, University of Birmingham,
Birmingham B15 2TT, United Kingdom}
\address[GANIL]{GANIL, CEA/DSM-CNRS/IN2P3, BP 5027, F-14076 Caen cedex, France}
\address[IPN]{Institut de Physique Nucl\'eaire, IN2P3-CNRS,
F-91406 Orsay cedex, France}

\thanks[email]{Corresponding author: orr@caelav.in2p3.fr}
\thanks[thFSU]{Present address: FSU, Tallahassee, USA.}
\thanks[thLPC]{Present address: LPC, Caen, France.}
\thanks[thIPN]{Present address: IPN, Orsay, France.}
\thanks[thINFN]{Present address: INFN, Catania, Italy.}

\bigskip\bigskip

\begin{abstract}

A systematic study of high energy, one-neutron removal reactions on 23
neutron-rich, psd--shell nuclei ($Z=5-9, A=12-25$) has been carried out.
The longitudinal momentum distributions of the core fragments 
and corresponding single-neutron
removal cross sections are reported for reactions on a carbon target.  
Extended Glauber model calculations,
weighted by the spectroscopic factors obtained from 
shell model calculations, are compared to the experimental results. 
Conclusions are drawn regarding the
use of such reactions as a spectroscopic tool and spin-parity assignments are 
proposed for $^{15}$B, $^{17}$C, $^{19-21}$N, $^{21,23}$O, $^{23-25}$F.
The nature of the weakly bound systems 
$^{14}$B and  $^{15,17}$C is discussed.   
 

\end{abstract}

\end{frontmatter}


Fragment momentum distributions have long been recognised as signatures
of the large spatial extent of the valence nucleons in halo nuclei \cite{Orr97}.  
Recently measurements of   
one-nucleon removal\footnote{The term ``knockout'', which  
has been employed to refer to such reactions \cite{Nav98},  
is not adopted here as it has long been 
used for $(p,2p)$ and $(e,e'p)$
reactions, the description of which is very different from that of 
absorption and diffraction in one-nucleon removal.}
reactions on light targets have 
been proposed as a spectroscopic tool for high-energy radioactive beams
\cite{Nav98,Tos99}.   
This approach has arisen from the development of reaction calculations 
in which the strong absorption limit \cite{Huf81}
and core excited states are accounted for \cite{Tos99}.  
More specifically, the integrated cross 
sections are related to 
spectroscopic factors using an extended version \cite{Tos99,Tos00} 
of the spectator core model \cite{Hus85}, 
whilst the momentum distributions are
derived in the opaque limit of the Serber model \cite{Han96,Esb96}.  
To date, this 
approach has been applied to a 
few near dripline and halo nuclei \cite{Nav98,Aum00,Gui00,Nav99}.

In this Letter the results of an investigation of high-energy one-neutron removal
reactions over a broad range of light, neutron-rich psd-shell nuclei
are reported.  The goals of the work were twofold. Firstly, to explore the evolution
in structure, and the manner in which it is manifested in the core fragment
observables, from near stability to dripline and halo systems.
Secondly, for many of the near stable
nuclei the ground state structure is well established and,
consequently, it has been possible to test the validity of 
one-neutron removal reactions as a spectroscopic tool. 

In the following, measurements of
the core fragment longitudinal momentum distributions and integrated cross sections
resulting from reactions on a C target are presented.  Comparison is made 
for both observables to the results of extended Glauber type calculations 
incorporating second order noneikonal 
corrections to the JLM parameterisation of the optical potential \cite{Sau00}.  
In the case
of those systems with unknown, or poorly defined ground state structures, 
probable spin-parity assignments have been made.  
    
The secondary beams were produced via the fragmentation on a 
490 mg/cm$^2$ thick C target
of an intense ($\sim$1$\mu$Ae) 70~MeV/nucleon
$^{40}$Ar$^{17+}$ beam provided by the GANIL coupled cyclotron facility.  
The reaction products were collected and selected according to magnetic rigidity 
using the SISSI device 
coupled with the alpha-shaped beam analysis spectrometer.   A mean rigidity of 
2.880 Tm was selected to allow for the transmission of nuclei from $^{12}$B to
$^{25}$F with energies in the range of 43 -- 71~MeV/nucleon (Table 1).   
The energy spread
in the secondary beams, as defined by the spectrometer acceptances, was 
$\Delta$E/E=2\%.  

The measurements of the momentum distributions and one-neutron removal
cross sections were performed using the SPEG spectrometer \cite{SPEG}.  
Owing to the
large energy spread in the secondary beam, SPEG was operated in a dispersion
matched energy-loss mode \cite{Orr92} for which a resolution in the momentum
measurements of $\delta$p/p = 3.5$\times$10$^{-3}$ was obtained.  Importantly 
the large angular acceptances 
of the spectrometer 
(4$^{\circ}$ in the vertical and horizontal planes) provided for complete
collection of the core fragments, obviating any ambiguities in 
the integrated cross sections and longitudinal momentum distributions
that would arise from limited transverse momentum acceptances \cite{Rii93}.  
Furthermore, 
the broad momentum acceptance of the spectrometer ($\Delta$p/p=7\%) 
allowed the momentum distributions for
one-neutron removal on all the nuclei of interest to be obtained in a 
single setting (B$\rho_{SPEG}$=2.551~Tm).
A secondary C reaction target of thickness 170~mg/cm$^2$ was employed for the 
measurements described here (the results obtained with a Ta target will be reported
elsewhere \cite{Sau00}).  

Ion identification at the focal plane of SPEG was achieved using the 
energy loss derived from a gas ionisation chamber and the time-of-flight
between a thick plastic stopping detector and the cyclotron radio-frequency.  
Additional identification information was 
provided by the residual energy measurement furnished by the plastic detector
and the time-of-flight with respect to a thin-foil microchannel
plate detector located at the exit of the beam analysis spectrometer.
Two large area drift chambers straddling the focal plane of SPEG were employed 
to determine the angles of entry of
each ion and, consequently, allowed the focal plane position spectra to be 
reconstructed.  
The momentum of each particle was then derived from
the reconstructed focal plane position.  Calibration in momentum was achieved 
by removing the reaction target and steping the mixed secondary beam of 
known rigidity along the focal plane.  This procedure also facilitated a
determination of the efficiency across the focal plane for the collection
of the reaction products.

The intensities of the various components of the
secondary beam were measured in runs taken with the secondary reaction
target removed and the spectrometer set to the same rigidity as the beamline.
These were calibrated in terms of the primary beam current, which
was recorded continuously throughout the experiment using a 
non-interceptive beam monitor.
Checks were also provided by the counting rates in the microchannel at the 
exit of the beam analysis spectrometer and a second located 
just upstream of the secondary reaction target.
Typical secondary beam intensities ranged from $\sim$600
$^{15}$C/s to $\sim$1 $^{25}$F/s.


\begin{table}
\begin{center}
\begin{tabular}{cccccc}
\hline
$^A$Z & Energy  & FWHM$_{cm}$ &  $\sigma_{-1n}$ & $\sigma_{-1n}^{Glauber}$ & J$^\pi$\\
      & [MeV/nucleon] &   [MeV/c]           & [mb]            &
[mb] & \\
\hline
$^{12}$B & 67 & 142 $\pm$ 3.5 & 81 $\pm$ 5         & 91  & 1$^+$    \\
$^{13}$B & 57 & 135 $\pm$ 7   & 59 $\pm$ 4         & 62  & 3/2$^-$  \\
$^{14}$B & 50 & 56.5 $\pm$ 0.5& 153 $\pm$ 15       & 185 & 2$^-$    \\
         & 86 & 57 $\pm$ 2 $^a$& 48 $\pm$ 5 $^a$   &     &     \\
         & 59 & 55 $\pm$ 2 $^b$ & 176 $\pm$ 16 $^b$&     &     \\
$^{15}$B & 43 & 73 $\pm$ 2.5  & 108 $\pm$ 13       & 89  & 3/2$^-$ $^c$\\
\hline

$^{14}$C & 71 & 180  $\pm$ 5   &  65 $\pm$ 4  & 89  & 0$^+$\\
$^{15}$C & 62 & 63.5 $\pm$ 0.7 & 159 $\pm$ 15 & 168  & 1/2$^+$\\
         & 85 & 67   $\pm$ 3 $^a$& 33 $\pm$ 3 $^a$ &   & \\
$^{16}$C & 55 & 108 $\pm$ 2    &  65 $\pm$ 6  & 75  & 0$^+$\\
$^{17}$C & 49 & 111 $\pm$ 3    &  84 $\pm$ 9  & 71  & 3/2$^+$ $^{c,a}$\\
         & 84  & 145 $\pm$ 5 $^a$ &  26 $\pm$ 3 $^a$  &   & \\
         & 96.8  & 94 $\pm$ 19 $^d$ &  41 $\pm$ 4 $^d$  &   & \\
         & 904  & 141 $\pm$ 6 $^e$ &  129$\pm$22 $^f$   &   & \\
$^{18}$C & 43 & 126 $\pm$ 5    & 115 $\pm$ 18 & 119  & 0$^+$ \\
         & 86.2  & 110 $\pm$ 12 $^d$    & 35 $\pm$ 2 $^d$ &   & \\

\hline
$^{17}$N & 65 & 141 $\pm$ 4 &55  $\pm$ 5   & 67  & 1/2$^-$ \\
$^{18}$N & 59 & 168 $\pm$ 3 &109 $\pm$ 11  & 91  & 1$^-$ \\
$^{19}$N & 53 & 177 $\pm$ 3 & 86 $\pm$ 9   & 83  & 1/2$^-$ $^{c,g}$\\
$^{20}$N & 48 & 162 $\pm$ 4 & 98 $\pm$ 13  & 101 & 2$^-$ $^{c}$ \\
$^{21}$N & 43 & 149 $\pm$ 7 &140 $\pm$ 44  & 151 & 1/2$^-$ $^c$ \\

\hline
$^{19}$O & 68 &190 $\pm$ 8   &104 $\pm$ 12 & 80   & 5/2$^+$ \\
$^{20}$O & 62 &219 $\pm$ 5   &112 $\pm$ 11 & 96   & 0$^+$ \\
$^{21}$O & 56 &210 $\pm$ 6   &134 $\pm$ 14 & 123  & 5/2$^+$ $^{c,g}$ \\
$^{22}$O & 51 &206 $\pm$ 4   &120 $\pm$ 14 & 140  & 0$^+$ \\
$^{23}$O & 47 &114 $\pm$ 9   & - $^k$          & 122  & 1/2$^+$ $^c$ \\
\hline

\end{tabular}
\end{center}
\end{table}

\begin{table}
\begin{center}
\begin{tabular}{cccccc}
\hline
$^A$Z & Energy  & FWHM$_{cm}$ &  $\sigma_{-1n}$ & $\sigma_{-1n}^{Glauber}$ & J$^\pi$\\
      & [MeV/nucleon] &   [MeV/c]           & [mb]            &
[mb] & \\
\hline

$^{22}$F & 64 & 185 $\pm$ 14 & 121 $\pm$ 16 & 61   & 4$^+$ \\
$^{23}$F & 59 & 235 $\pm$ 4  & 114 $\pm$ 12 & 106  & 5/2$^+$ $^{c,h,i}$\\
$^{24}$F & 54 & 129 $\pm$ 4  & 124 $\pm$ 16 & 109  & 1$^+$ $^{c}$, 3$^+$ $^{c,j}$ \\
$^{25}$F & 50 & 106 $\pm$ 8  & 173 $\pm$ 46 & 154  & 5/2$^+$ $^c$ \\
\hline \\

\multicolumn{6}{l}{\footnotesize
$a$ ref. \protect\cite{Baz98} (Be target), $b$ ref. \protect\cite{Gui00} 
(Be target)} \\
\multicolumn{6}{l}{\footnotesize
$c$ assignment from present experiment, $d$ ref. \protect\cite{Baz95} 
(Be target)} \\
\multicolumn{6}{l}{\footnotesize
$e$ ref. \protect\cite{Bau98} (C target), $f$ ref. \protect\cite{ENPE99} 
(C target), $g$ ref. \protect\cite{Cat89}, 
$h$ ref. \protect\cite{Orr89}, $i$ ref. \protect\cite{Goo74} } \\
\multicolumn{6}{l}{\footnotesize
$j$ ref. \protect\cite{Ree99}, $k$ no beam intensity normalisation available} \\ \\

\end{tabular}

\caption{Summary of results for one-neutron removal.}

\end{center}
\end{table}


The longitudinal momentum distributions 
for the core fragments arising 
from one-neutron removal
are displayed in figure 1 and the extracted widths (FWHM in the projectile frame) 
are summarised in Table 1.  
The widths were
derived from Gaussian fits to the central regions of 
each distribution.  The effects arising from the target (straggling etc), efficiency
along the focal plane and 
instrumental resolution have been taken into account in deriving the final values.  
The corresponding one-neutron removal cross
sections are listed in Table 1 and displayed in figure 2.
The uncertainties quoted include the contributions from both the
statistical uncertainty and that arising from the determination of the 
secondary beam intensity ($\sim$7\%).

A number of features are immediately apparent on inspection of figures 1 and 2.
Firstly, the crossing of the N=8 shell and N=14 
sub-shell closures
are associated with a marked reduction in the widths of the core momentum 
distributions (viz, $^{14,15}$B, $^{15}$C, $^{23}$O and $^{24,25}$F).  Secondly, 
with respect to the neighbouring isotopes,
$^{14}$B and $^{15}$C exhibit enhanced one-neutron removal cross sections.
The former effects arise from the large $\nu$2s$_{1/2}$ admixtures expected in the ground
states of
the Z=4-6, N=9 isotones \cite{Ren97} (see below), which may also persist for N=10, as
suggested by recent studies of $^{14}$Be \cite{Suz99,Lab00}.
A narrowing of the momentum distributions may also be expected for N=15 and 16
as in a simple shell model picture the valence neutrons occupy the $\nu$2s$_{1/2}$ 
orbital.
In general terms, the enhanced cross sections may be attributed to a combination 
of weak binding ($^{14}$B: S$_n$= 0.97 MeV; 
$^{15}$C: S$_n$= 1.22 MeV) and the 
large $\nu$2s$_{1/2}$ admixtures in the ground state wavefunctions, which 
may be related to
extended valence neutron density distributions, as discused below.

As noted in Table 1, the present measurements may be compared to those
made for $^{14}$B \cite{Gui00,Baz95} and 
$^{15,17,18}$C \cite{Baz95,Baz98,Bau98}.  While
agreement is found for the momentum distributions, the integrated cross
sections are systematically some 3-5 times higher than those
reported at similar energies using the A1200 fragment 
separator \cite{Baz95,Baz98}.  Analysis of the 
transverse momentum distributions obtained in the present experiment
demonstrate that the rather limited acceptances of the A1200 are the origin of this 
discrepancy \cite{Sau00}.  In the case of $^{14}$B, the present results and
those of ref. \cite{Gui00}, also obtained using a high acceptance spectrometer,
are in good accord.

In order to make a more quantitative analysis of the measurements and examine the 
utility of such reactions as a spectroscopic
tool, extended Glauber type calculations have been carried out.   
The calculations, the principal features of which follow 
refs. \cite{Tos99,Hen96,Neg99}\footnote{A similar spectator core description and
treatment of core excited states was developed earlier by  
Sagawa {\em et al.} in a study of inclusive momentum distributions following
neutron removal on $^{11}$Be \cite{Sag90}.}, include absorption (or 
stripping) and
diffractive (or elastic) one-nucleon breakup.  An important feature is that
the S-matrices describing 
these processes have been derived from the microscopic 
interaction of Jeukenne, Lejeune and
Mahaux (JLM) \cite{JLM} within an eikonal approximation 
employing noneikonal corrections \cite{Wal73,Car93}. 
As discussed 
by Bonaccorso and Carstoiu \cite{Bon00} and Tostevin
\cite{Tos00}, such microscopic potentials are much better adapted to the intermediate
energy range than optical limit \cite{Tos99} or global parameterisations \cite{Hen96}.
In addition to the cross sections, 
the core longitudinal momentum distributions 
have been
computed within this framework, as opposed to the black disk 
approximation of ref. \cite{Han96}.  
A detailed description
of the calculations, together with the results obtained for the transverse
momentum distributions and with a Ta target, will be presented elsewhere \cite{Sau00}.

In terms of structure, overlaps were calculated between the ground state 
wavefunctions of the projectiles
($J^{\pi}$) and the core states 
($I^{\pi}_{c}$) coupled
to a valence neutron ($nlj$).  The single-particle wavefunctions were defined within
a Woods-Saxon potential with fixed geometry ($r_0$=1.15~fm, $a_0$=0.5~fm for Z=5 and 6; 
$r_0$=1.2~fm, $a_0$=0.6~fm for Z=7-9) with the depth adjusted to reproduce the 
effective binding energy ($S_n^{eff}$) which was fixed as the sum of the 
single-neutron separation energy and the excitation energy of the core state.
The cross section to populate a given core final state is then,

\begin{equation}
\sigma(I^{\pi}_{c}) \ = \ \sum_{nlj} C^2S(I^{\pi}_{c},nlj) \sigma_{sp}(nlj,S_n^{eff}) 
\end{equation}

where $C^2S$ is the spectroscopic factor for the removed neutron with respect 
to the core
state and $\sigma_{sp}$ is the cross section for removal of the neutron by 
absorption ($\sigma_{abs}$), 
diffraction ($\sigma_{diff}$) and Coulomb dissociation 
(only $\sim$7~mb in the most favourable cases -- $^{14}$B and $^{15}$C \cite{Sau00}).
The total inclusive one-neutron removal cross section ($\sigma_{-1n}^{Glauber}$) 
is then the sum over the cross sections
to all core states.  Similarly, the inclusive core momentum distribution
is the sum of all core state momentum distributions, weighted by the corresponding
cross sections. 
Within the framework of the spectator core description used here,
excitation of the core in the reaction and final-state interactions are neglected.  

The spectroscopic factors employed here have been calculated with the shell model
code OXBASH \cite{OXBASH} using the WBP interaction \cite{WBP} within the 
1p-2s1d configuration space.
Where known, the experimentally established spin-parity assignments and 
core excitation
energies have been used.  In all other cases the shell model predictions 
have been assumed.  
The resulting cross sections and momentum distributions are displayed in
Table 1 and figures 1 and 2.  The breakdown of the calculated cross sections over
the core states for each nucleus is detailed in ref. \cite{Sau00}; 
as examples, and to aid in the following discussion, the results
are listed for $^{14}$B and  $^{15,17}$C in Table 2.   As the momentum 
distributions reflect the orbital angular momentum of the removed neutron, the 
calculated distributions have been normalised to the peak number
of counts to facilitate the comparison (figure 1).  For all the nuclei observed,
including those with known structure, very good agreement
is found between the calculated and measured distributions and cross
sections, with the exception of $^{22}$F, where the cross section
is underestimated.  Consequently, spin-parity
assignments, derived from the shell model predictions, have been 
proposed for $^{15}$B, $^{17}$C, $^{19-21}$N, $^{21,23}$O, $^{23-25}$F (Table 1).
In the case of $^{24}$F, a 3$^+$ or 1$^+$ assignment appears possible based 
on the present data \cite{Sau00}.  
The decay study of Reed {\em et al.} suggests, however, that the
former is the most likely \cite{Ree99}, in line with the shell model predictions.


\begin{table}
\begin{center}
\begin{tabular}{c c c c c c c}

\hline
\multicolumn{1}{l}{$^{14}$B}  J$^\pi_{gs}=2^-$ & & & & & & \\
\hline
 E$_{x}$($^{13}$B) [MeV] & I$^\pi_c$ & $nlj$  & $C^2S$ & $\sigma_{abs}$ [mb] & 
 $\sigma_{diff}$ [mb] & $\sigma(I^{\pi}_c)$ [mb]\\
\hline
 0.0 & $3/2^-$ & 1d$_{5/2}$ & 0.31 & 9.6 & 8.3 & 18.3 \\
      &         & 2s$_{1/2}$ & 0.64 & 57. & 59. & 121.7 \\
3.483 & $3/2^+$ & 1p$_{1/2}$ & 0.41 & 8.5 & 6.8 & 15.6 \\
3.68  & $5/2^+$ & 1p$_{1/2}$ & 0.8  & 16.2& 12.9& 29.5 \\
\hline
      &          &           &      &     & \multicolumn{2}{c}{$\sigma_{-1n}^{Glauber}$=185 mb}\\
\hline

\multicolumn{1}{l}{$^{15}$C}  J$^\pi_{gs}=1/2^+$ & & & & & & \\
\hline
  E$_{x}$($^{14}$C) [MeV] & I$^\pi_c$ & $nlj$  & $C^2S$ & $\sigma_{abs}$ [mb] & 
  $\sigma_{diff}$ [mb] & $\sigma(I^{\pi}_c)$ [mb]\\
\hline
 0.0 & $0^+$ & 2s$_{1/2}$ & 0.83  & 62.1 & 55.8 & 124.2\\
6.094 & $1^-$ & 1p$_{3/2}$ & 0.16  & 2.8  & 1.9  & 4.7  \\
      &       & 1p$_{1/2}$ & 1.03  & 16.3 & 10.6 & 27.2 \\
6.903 & $0^-$ & 1p$_{1/2}$ & 0.46  & 6.9  & 4.4  & 11.5 \\
\hline
      &          &           &      &     & \multicolumn{2}{c}{$\sigma_{-1n}^{Glauber}$=168 mb}\\
\hline

\multicolumn{1}{l}{$^{17}$C}  J$^\pi_{gs}=3/2^+$ & & & & & & \\
\hline
  E$_{x}$($^{16}$C) [MeV] & I$^\pi_c$ & $nlj$  & $C^2S$ & $\sigma_{abs}$ [mb] & 
  $\sigma_{diff}$ [mb] & $\sigma(I^{\pi}_c)$  [mb]\\
\hline
 0.0 & $0^+$ & 1d$_{3/2}$  &  0.035 & 0.9 & 0.8 & 1.7 \\
 1.762 & $2^+$ & 1d$_{5/2}$ &  1.41  & 29.3& 24.7& 54.8 \\
       &       & 2s$_{1/2}$ &  0.16  & 6.9 & 6.8 & 14.1 \\
\hline
      &          &           &      &     & \multicolumn{2}{c}{$\sigma_{-1n}^{Glauber}$=71 mb}\\
\hline  \\

\end{tabular}
\end{center}
\caption{Calculated one-neutron removal cross sections for $^{14}$B and $^{15,17}$C.}

\end{table}


Of particular interest amongst the nuclei investigated here are 
$^{14}$B and  $^{15,17}$C, which, based on the relatively weak binding of the 
valence neutrons and measurements of the core momentum distributions and
one-neutron removal cross sections,
have been suggested to be one-neutron halo 
systems \cite{Gui00,Baz95,Baz98}.  
As may be seen in figures 1 and 2 and Table 2, the momentum distributions and cross sections
for $^{14}$B and  $^{15}$C are well reproduced by the present calculations
employing the spectroscopic factors derived from the shell model,
in which the ground state wavefunctions are predominately a 2s$_{1/2}$ valence neutron 
coupled to the core ($^{13}$B and  $^{14}$C) in the ground state, as suggested 
by decay studies
\cite{Alb74} and single neutron-transfer experiments \cite{AJZ86}.  In the case of $^{17}$C, a 
spin-parity assignement of 3/2$^{+}$ is favoured, whereby the ground state 
configuration is
predominately a 1d$_{5/2}$ valence neutron coupled to the $^{16}$C core 2$^+_1$
state.  This confirms the suggestion of Bazin {\em et al.} \cite{Baz98} and the 
calculations of Ren {\em et al.} \cite{Ren96}, and is
supported by the recent observation of the 1.76~MeV gamma-rays de-exciting the 2$^+_1$
state in $^{16}$C following one-neutron removal on $^{17}$C \cite{PGH00}.
Such a structure, with a high $S_n^{eff}$ (2.49 MeV) and a valence neutron
angular momentum of {\em l}=2, excludes the possibility of any halo structure 
developing
as evidenced by measurements of the total reaction cross section 
\cite{Vil89,StL89,Vil91,Oza98}.

Moderate enhancements, however, have been observed in the total reaction cross
section measurements for $^{14}$B \cite{Vil89,StL89,Vil91}.  
Together with the  
ground state structure deduced from the present experiment and refs. \cite{Gui00,Baz98},
it seems probable that a spatially extended valence neutron density distribution 
does occur; although the one-neutron binding energy of nearly 1~MeV
will supress the development of a distribution as large as that found in the
more weakly bound one-neutron halo nuclei $^{11}$Be and $^{19}$C.
Detailed measurements of the total reaction cross section over a range of 
energies would thus be of particular interest in mapping out the
density distribution of $^{14}$B. 

In the case of $^{15}$C the situation is unclear, with measurements
of the total reaction cross section exhibiting no effect \cite{Vil89,StL89,Oza98} 
and small enhancements \cite{Vil91,Fan00}.  Despite the predominately 
${\em l}$=0 character 
of the valence neutron, the higher neutron binding energy of $^{15}$C 
(S$_n$=1.22~MeV) should restrict further the spatial extent of the neutron density 
distribution.  Interestingly, very recent measurements of the 
charge-changing cross sections 
for the C isotopes exhibit an increase for $^{15}$C \cite{Chu00}.  

In summary, a systematic investigation of one-neutron removal
reactions has been carried out on a series of
neutron-rich psd-shell nuclei.  The longitudinal momentum distributions
and corresponding single-neutron removal cross sections for the core fragments 
were measured using a high acceptance spectrometer.
Extended Glauber model calculations, coupled with spectroscopic factors derived
from shell model calculations employing the WBP interaction,
reproduce well the momentum distributions and cross sections. On this basis 
spin-parity assignments
have been proposed for $^{15}$B, $^{17}$C, $^{19-21}$N, $^{21,23}$O, $^{23-25}$F.
Given the ground state configurations deduced here
and measurements of the total reaction cross section,  
it is suggested that $^{14}$B presents a moderately extended valence neutron
density distribution.  This does not 
appear to be the case for $^{17}$C, whilst $^{15}$C exhibits  
contradictory behaviour.

In more general terms it is concluded that high energy one-nucleon removal reactions
represent a powerful spectroscopic tool far from stability.  
Moreover it has been demonstrated that coupled with a high acceptance, broad range
spectrograph, such reactions offer a means to survey structural evolution
over a wide range of isospin in a single experiment.
The current development
of large area, highly segmented, multi-element Ge-arrays \cite{Mue00,EXOGAM} 
should further enhance the
sensitivity of such studies.

{\bf Acknowledgements}

The support provided by the staffs of {\sc lpc} and {\sc ganil} is gratefully
acknowledged. Discussions with J.A.~Tostevin are also
acknowledged, as is the guidance provided by B.A. Brown into the intricacies
of the shell model and the assistance provided by G.~Mart\'inez in 
preparing the experiment. This work was funded under the auspices 
of the {\sc in2p3-cnrs}
(France) and {\sc epsrc} (United Kingdom). Additional support from the Human 
Capital and Mobility Programme of the European Community
(contract n$^\circ$ CHGE-CT94-0056) and the GDR Noyaux Exotiques ({\sc cnrs}) 
is also acknowledged.

\newpage

\begin{figure}[h]
\begin{center}
\mbox{\epsfig{file=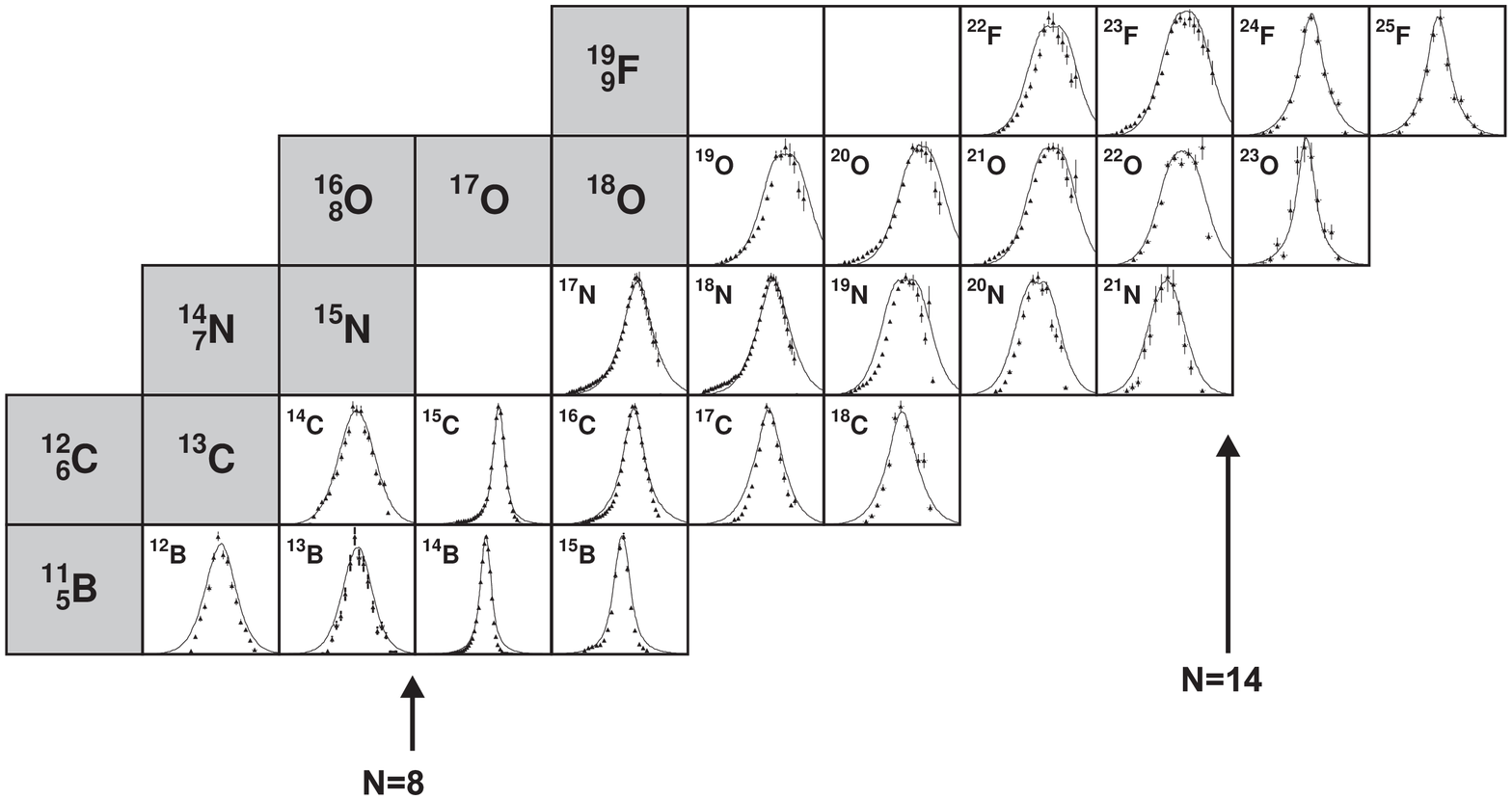,height=10cm,angle=90}}
\end{center}
\caption{Core fragment longitudinal momentum distributions for one-neutron removal on C. 
The solid lines correspond to the Glauber model calculations (see text for details).}
\end{figure}

\begin{figure}[h]
\begin{center}
\mbox{\epsfig{file=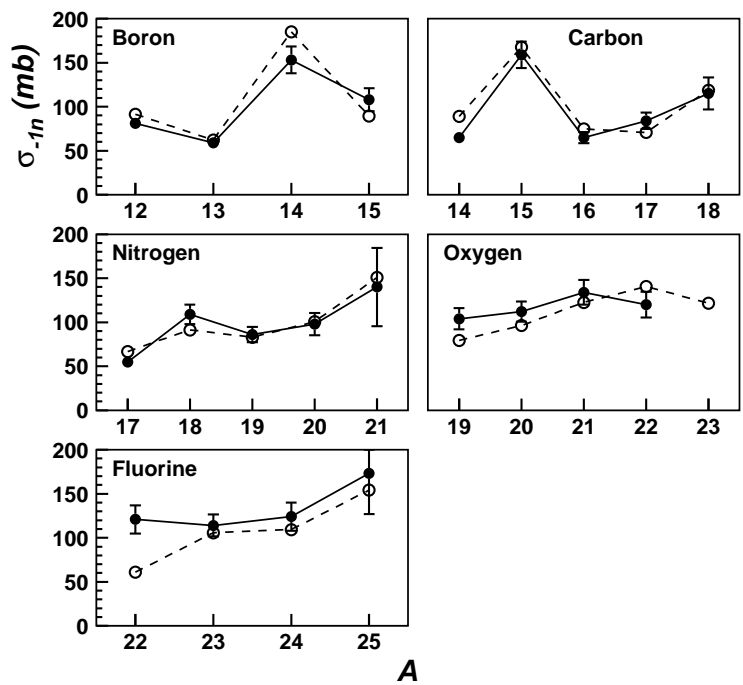,height=10.5cm}}
\end{center}
\caption{Measured (solid points and line) and calculated (open points and dotted line) 
one-neutron removal cross sections.}
\end{figure}

\end{document}